\documentclass[preprint,NumberedRefs]{JASAnew}
\usepackage[tight,footnotesize]{subfigure}

\begin{document}
%% the square bracket argument will send term to running head in
%% preprint, or running foot in reprint style.

\title[]{Subspace-based compressive sensing algorithm for raypath separation in a shallow-water waveguide}

% ie
%\title[JASA/Sample JASA Article]{Sample JASA Article}

%% repeat as needed
\author{Longyu Jiang}

% ie
%\author{Author Two}
%\author{Author Three}

\affiliation{The Laboratory of Image Science and Technology, Southeast University, Nanjing
210096, China.}
% ie
%\affiliation{Acoustic Science and Technology Laboratory, Harbin Engineering University, Harbin 150001,China.}

 \thanks{Also at Acoustic Science and Technology Laboratory, Harbin Engineering University, Harbin 150001,China.}
 
\author{Zhe Zhang}
%\author{Author Two}
%\author{Author Three}

\affiliation{The Laboratory of Image Science and Technology, Southeast University, Nanjing
210096, China.}

\author{Rui Jin}
%\author{Author Two}
%\author{Author Three}

\affiliation{The Laboratory of Image Science and Technology, Southeast University, Nanjing
210096, China.}

\author{Xiao Zhou}
%\author{Author Two}
%\author{Author Three}

\affiliation{The Laboratory of Image Science and Technology, Southeast University, Nanjing
210096, China.}
%% for corresponding author
\email{JLY@seu.edu.cn}
%% for additional information
\thanks{}

\author{Philippe Roux}

\affiliation{Institut des Sciences de la Terre, Universit\'{e} Joseph Fourier,  Centre National de la Recherche Scientifique, 1381 Rue de la Piscine, Saint-Martin d'H\`{e}res, France.}
% ie
% \author{Author Four}
% \email{author.four@university.edu}
% \thanks{Also at Another University, City, State ZipCode, Country.}

%% For preprint only,
%  optional, if you want want this message to appear in upper left corner of title page
% \preprint{}

%ie
%\preprint{Author, JASA}		

% optional, if desired:
\date{\today} 

\begin{abstract}
% Put your abstract here. Abstracts are limited to 200 words for
% regular articles and 100 words for Letters to the Editor. Please no
% personal pronouns, also please do not use the words ``new'' and/or
% ``novel'' in the abstract. An article usually includes an abstract, a
% concise summary of the work covered at length in the main body of the
% article.   
Compressive sensing (CS) has been applied to estimate the direction of arrival (DOA) in underwater acoustics. However, the key problem needed to be resolved in a {multipath} propagation environment is to suppress the interferences between the raypaths. Thus, in this paper, {a subspace-based compressive sensing algorithm that formulates the statistic information of the signal subspace in a CS framework is proposed.}
The experiment results show that (1) the proposed algorithm enables the separation of raypaths that arrive closely at the {receiver} array and (2) the existing algorithms fail, especially in a low signal-to-noise ratio (SNR) environment.  	
\end{abstract}

%% pacs numbers not used
\maketitle

\section{Introduction}
\setlength{\parindent}{5ex}

\medskip

%The problem of target detection and location in shallow sea is a basic one of marine scientific research and technical application. As there are reflections or refractions during the propagation of acoustic signals in the surface and the seabed, the acoustic signals emitted by source will be received by sensors in multi paths which can be utilized to determine the position of the source. However, the weakening of the signal amplitude and the interference between the signals have affected the detection and positioning of the sources in the shallow water multi-path propagation. Therefore array signal processing was raised to suppress the interference to realize the separation and determine the positions of the sources. 

{Multipath} acoustic waves propagate in shallow water due to the reflection and (or) refraction of {acoustic eigenrays.}
%the water surface {\color{blue}or the continental shelf}.
 {In ocean acoustic tomography(OAT)\cite{Munk1995}, we use the variation in the arrival time of rays to inversely derive fluctuations of sound speed profile in the water column.
 Recent works have emphasized the use of angle variations to image sound speed perturbations using the sensitivity kernel approach \cite{AulanierShallow2013,AulanierTime2013}. Multiple paths cover the different parts of the ocean, so they can provide spatially-extended information for the inversion process. However, they also produce interferences between them. Thus, successful separation of these raypaths based on direction-of-arrival (DOA) estimation is the first step of OAT. Researchers have paid great attention to solving this problem, and many algorithms have been proposed\cite{schimidt1986,jiang2017raypath,jiang2017Active}.}
%The estimation of Direction of Arrival (DOA) is a key problem in the field of array signal processing, and it has been widely applied in radar, communication, earthquake and other fields. 
Conventional beamforming (CBF) and Capon BF are two such conventional algorithms, but they fail in separating raypaths that {arrive close to each other at the receiver array}. 
%These algorithms greatly improved the resolution compared to CBF.  
{Moreover, Iturbe and co-authors proposed a double-beamforming algorithm in a double-array configuration to improve the performance of the conventional beamforming algorithm \cite{Iturbe2008}. On the other hand, high-resolution algorithms are also introduced to improve the angular resolution of beamforming.
 {The MUltiple SIgnal Classification (MUSIC) algorithm \cite{schimidt1986} is a representative of the high-resolution algorithms, which is based on the orthogonality of the signal and noise subspace.} Following the strategy of high-resolution algorithms, exploring the spectral information of the emitted signal, Jiang et al proposed a smoothing-MUSICAL algorithm \cite{jiang2017raypath} to separate the correlated or coherent raypaths and a higher-order algorithm\cite{jiang2017Active} in the frequency domain to enable the separation the raypaths in an ocean environment in the presence of colored noise.
 Experimental results illustrate that both algorithms achieve resolution improvement compared to beamforming-like algorithms.}
%To further improve their resolutions, the MUltiple SIgnal Classification (MUSIC) algorithm \cite{schimidt1986} has been developed {to decompose the covariance matrix of the received signals into a signal subspace and a noise subspace and to improve the performance of BF-like algorithms, mainly using their orthogonality.} 
% Quality control editor: It is unclear what "their resolutions" refers to in the previous sentence. Please consider replacing it with "the resolutions of these approaches" or a more appropriate phrase.
However, {the noise and signal subspaces} are not always totally orthogonal in actual operation\cite{thomas1995}. Thus, the DOA estimation based on subspace decomposition still faces great challenges,{especially when multiple raypaths propagate in a strongly noisy ocean environment}. Recently, the DOA estimation based on CS has been {very} attractive {to many researchers in this area}. 
CS theory as a new information technology makes use of the fact that {received} signals with inherent sparsity can be reconstructed perfectly from very few measurements {through the solution of} a convex minimization problem.
%\cite{donoho2006compressed}. 
%It has been applied to solving many scientific issues in different domains, such as medical imaging \cite{Candes2006Robust}, aeroacoustic imaging \cite{zhong2013compressive} et etc.
In underwater acoustics, Edelmann and Gaumond \cite{Edelmann2011Beamforming} first applied a {CS} algorithm to estimate DOAs using ship-towed horizontal arrays. Compared with conventional beamforming, the algorithm achieved a resolution improvement and interference suppression. Xenaki and Gerstoft \cite{Xenaki2014Compressive} developed a compressive beamforming algorithm and further discussed its performance, especially with coherent arrivals, single-snapshot data and random array geometries. In addition, the authors indicated that the limitations of CS-based algorithms are related to the beampattern. Gerstoft et al. \cite{gerstoft2015multiple} extended the algorithm to the case of multiple snapshots combining a maximum a posteriori method with the least absolute shrinkage and selection operator (LASSO) method. The experiment shows that this method also outperforms conventional high-resolution methods even under challenging scenarios such as coherent arrivals and single-snapshot data. {Recently, a compressive sensing beamforming method (CSB-II) \cite{zhong2013compressive} is developed based on sampling covariance matrix in the application context of aeroacoustic experiments containing strong background noise at broadband frequencies. The experiment results illustrate that the CSB-II method achieves a higher resolution and is robust to sensing noise for presumably spatially sparse and incoherent signals. However, 
 in the context of ocean acoustic tomography, the multiple raypaths are produced by the emitted signal through the refraction or the reflection of the ocean surface and the continental shelf. {Thus, they are always fully correlated or coherent. The scattering at different points on the sea surface or the eigenpaths crossing different areas of the water column may attenuate these correlations; however, there is still the possibility that the case of fully correlated or coherent raypaths exists. This is a very difficult separation case.
 % For the case, the signal subspace would be of dimension 1, so we use the frequency-smoothing as a preprocessing. 
 In addition, even in the case of partly correlated raypaths, the correlations between the raypaths impede the successful separation or decrease the separation resolution of raypaths. However, the existing CS-based algorithms do not focus on solving this problem.} 
 
In order to perform the separation between eigenrays, the present paper proposes {a subspace-based compressive sensing algorithm that formulates the statistic information of the signal subspace in a CS framework. The signal subspace is first obtained through an eigenvalues decomposition to the spectral matrix of the received signal. Then, to suppress the interferences produced by the correlations between the raypaths, 
% which is the key problem needed to be resolved in a {multipath} propagation environment,   
 the sparse signal reconstruction is achieved by designing a subspace-based {$L_1$} regularization optimization. }}
 
 {The paper is organized as follows. In Section II, the signal model is established. In Section III, the principle of the comparative method--reweighed compressive sensing is described. In Section IV, we introduce our proposed algorithm. In Section V, we illustrate the performance of the proposed algorithm based on simulations, a tank experiment and an ocean experiment.}

\section{Signal model}

In a shallow-water waveguide, we assume that  {$P$ raypaths emitted by a far-field source $e(t)$ are received by each element of a vertical array composed of $M$ sensors.} $s_p (t)=a_p e(t)$ denotes the $p^{th}$ raypath, where $a_p$ is the random amplitude of the $p^{th}$ path. The received signal on the $m^{th}~(m=1,2,...,M)$ sensor, denoted by $y_m(t)$, can be modeled in the time domain as follows.
%\begin{equation}
%	{y_m}(t) = \sum\limits_{p = 1}^P {{g_p}({\theta _p}){s_p}(t - {\tau _{mp}}) + {n_m}(t)} 
%\end{equation}

{\begin{equation}
	{y_m}(t) = \sum\limits_{p = 1}^P {{s_p}(t - {\tau _{mp}}) + {n_m}(t)} 
\end{equation}}
{where $s_p$ denotes the amplitude of the $p^{th}$ source on the $m^{th}$ sensor}; and $n_m(t)$ is the white Gaussian noise received at the $m^{th}$ sensor. $\tau_{mp}$ represents the propagation delay of the $p^{th}$ raypath between the $m^{th}$ sensor and the reference sensor. {}

In {the} frequency domain, Eq. 1 can be rewritten as
%\begin{equation}
%	{Y_m}(\nu) = \sum\limits_{p = 1}^P {{S_p}(\nu)\exp ( - j\nu (m - 1){\Phi _p}) + {N_m}(\nu)} 
%\end{equation}
{\begin{equation}
	{Y_m}(\nu) = \sum\limits_{p = 1}^P {{S_p}(\nu)G_{mp} + {N_m}(\nu)} 
\end{equation}}
with $G_{mp}=exp(-j \nu (m-1) 2 \pi \frac{d sin \theta_p}{c})$, {$d$ is the distance between two adjacent sensors and $c$ is the propagation velocity of the acoustic signal, which is assumed to be constant in this paper. }{The term
%of the $p^{th}$ raypath on the $m^{th}$ sensor at frequency $\nu$. 
$S_p(\nu)=a_p e_p (\nu)$, where $a_p$ is the random amplitude of the $p^{th}$ path and $e_p(\nu)$ is the deterministic Fourier Transform of the emitted signal $e(t)$.} ${N_m}(\nu)$ is the received white Gaussian noise at frequency $\nu$. and $\theta_p$ is the direction of arrival of the $p^{th}$ raypath at the reference sensor.
% In Equation (3), 

%{\color{blue}where the term ${S_p}(\nu)$ is the deterministic amplitude of the $p^{th}$ raypath on the $m^{th}$ sensor at frequency $\nu$.}

Considering the received signals of all the $M$ sensors and constructing the sparse signal, 
%we have
%\begin{equation}
%	\begin{bmatrix}
%	Y_1\\ 
%	\vdots \\ 
%	Y_i\\ 
%	\vdots \\ 
%	Y_M
%	\end{bmatrix}
%	=\begin{bmatrix}
%	G_{11} & \cdots  & G_{1k} &  \cdots &G_{1P} \\ 
%	\vdots &  & \vdots &  &\vdots \\ 
%	G_{i1}&\cdots  &G_{ik}  &\cdots  & G_{iP}\\ 
%	\vdots &  &\vdots  &  &\vdots \\ 
%	G_{M1} &\cdots  &G_{Mk}  &\cdots  & G_{MP} 
%	\end{bmatrix}
%	\begin{bmatrix}
%	S_1\\ 
%	\vdots\\ 
%	S_i\\ 
%	\vdots\\ 
%	S_P
%	\end{bmatrix}
%	+\begin{bmatrix}
%	N_1\\ 
%	\vdots\\ 
%	N_i\\ 
%	\vdots\\ 
%	N_M
%	\end{bmatrix}
%\end{equation}
%Where $Y_i$ denotes the observation on the $i^{th}$ sensor. $G_{ik}$ is the steering vector of the $i^{th}$ sensor with the $k^{th}$ signal. $N_i$ is the noise on the $i^{th}$ sensor.
 Eq. 2 can be rewritten in matrix form as follows.
\begin{equation}
	{\bf Y} = \bf{G}{\mathbf S} + {\bf{N}}
\end{equation}
%Where $\mathbf{G}\in \mathbb{R}^{M\times N}$ is the array manifold. We assume that $\mathbf{S}\in \mathbb{R}^{N\times 1}$ and $\mathbf{N}\in \mathbb{R}^{M\times 1}$ are zero-mean.
where $\mathbf Y (\nu)=[Y_1(\nu),Y_2(\nu),\ldots, Y_M(\nu)]^{T}$ and $\mathbf Y(\nu)$ represents the received signal on all the $M$ sensors, whose size is $M \times 1$.
$\mathbf N(\nu)=[ n_1(\nu), n_2(\nu),\ldots, n_M(\nu)]^{T}$ is a vector of dimension $M \times 1$, which is obtained by the concatenation of the noise at each sensor. $\mathbf G=[\mathbf{G}_1,\mathbf{G}_2,\ldots,\mathbf{G}_M]^{T}$, and $\mathbf{G}_m=[e^{-2i\pi \nu \tau_{m1}},\ldots,e^{-2i\pi \nu \tau_{mQ}}]$, which is {an} over-complete dictionary. $\mathbf G$ contains the terms $e^{-2i\pi\nu \tau_{mp}}, p=1, 2, \ldots, P$, which describe the transfer functions between the source and the sensors. $\mathbf S =[S_1, S_2, \cdots, S_P, \cdots, S_Q]^{T}$ denotes the sparse-signal vector containing the angle and the amplitude information of the raypaths, which are produced from the emitted signal. {$Q$}, (${P<M}, P<<Q$ and $M<Q$) is the dimension of the sparse-signal vector, and $T$ means ``transposed''. {The so-called sparsity means that the number of nonzero entries in $\bf S $ is pretty small, i.e., ${\parallel \bf S \parallel}_{0}<Q$.}

{\section{Reweighed compressive sensing}

{Based on the signal model in frequency domain described above, we first review the compressive sensing beamforming algorithm.} {The signal $\bf S$ is sparse, as it has only $P$ nonzero components, with $P << Q$.} Thus, the seemingly underdetermined problem is considered. {{The sparse vector estimate $\mathbf S$ contains the angle and amplitude information of the raypaths,}}
%The angle and amplitude information of the $P<M$ raypaths is comprised of the sparsity solution $\mathbf S$, 
and it can be estimated by the following $l_1$ minimization problem:

%\begin{equation}
%\centering
%\hat{S}=min \|S\|_1 \ \ \  subject \ \  to \ \ \  \mathbf G S=Y
%\end{equation}
%where $\hat{S}$ is the estimation of the unknown $S$.

\begin{equation}
\centering
\min_{\mathbf S \in \mathbb C^{M}}^{}\| \mathbf {S} \|_{1} \textstyle{\ subject \ to} \ \| \mathbf {GS}-\mathbf Y  \|_{2} \leq \epsilon
\end{equation}
{where the noise is assumed to be spatially white and $\varepsilon$ is an upper bound for the noise norm.}
, such that $ \| \mathbf N \|_{2} \leq \epsilon$.

The reweighed CS \cite{Xenaki2014Compressive} improves the performance of the conventional CS algorithm through {the use of a diagonal weight matrix ${\bf W}_{nn} \in {\mathbb R}$ to enhance the sparsity of the solution:}

\begin{equation}
\centering
\min_{\mathbf S \in \mathbb C^{M}}^{}\| \mathbf {WS} \|_{1} \textstyle{\ subject \ to} \ \| \mathbf {GS}-\mathbf Y  \|_{2} \leq \epsilon
\end{equation}
%where $\epsilon$ is an upper bound of the noise. Even if $\epsilon$ is set to be lower than the power of the noise, the estimate is still accurate. 
The weight matrix, $\mathbf W= \textstyle {diag} [w_1, w_2, \cdots, w_Q]$, is initialized with the identity matrix $\mathbf I_Q$, and the elements $w_{n}^{k+1}$ at the $(k+1)^{th}$ iteration are updated as

\begin{equation}
\centering
w_{n}^{k+1}=\frac{1}{|\hat{s}_{n}^{k}|+{\xi}^{'}}
\end{equation}
where $ \widehat{\mathbf S}_{n}^{k}= \textstyle {diag} [\hat{s}_{n}^{1}, \hat{s}_{n}^{2}, \cdots, \hat{s}_{n}^{k}]$ is the solution in the previous iteration. Setting parameter $\xi>0$ ensures that a null element in the current estimate, $\widehat{\mathbf S}_{n}^{k}$, prevents instability in the iteration. The algorithm will proceed until solutions in adjacent iterations finally reach approximately equal values; i.e., $\widehat{\mathbf S}_{n}^{k+1}=\widehat{\mathbf S}_{n}^{k}$, where

\begin{equation}
\centering
|w_{n}^{k+1} {s}_{n}^{k+1}| =
\begin{cases}
\frac{s_{n}^{k}}{|s_{n}^{k}|+\xi} \approx 1 \ \  |s_{n}^{k}| >0   , \\
0, \ \  |s_{n}^{k}|=0 \\
\end{cases}
\end{equation}
which results in its convergence.}

{{In addition, {for stationary sources, which do not move in space, and for the whole $L$ snapshots,  which is a single observation vector of an array data, in the time domain}, due to the consistency of the sparsity the problem is solved by minimizing the $l_1$-norm of the product of the weight $\bf W$ and the vector ${\bf S}_{l_1}$ combined with the $l_1$-norm of the row vectors in $\bf S$,
\begin{equation}
{\min}{\left\|{\bf WS}_{l_1}\right\|_1}\text{ subject to }\left\| \bf GS-Y\right\|_2\leq\epsilon
\label{eq:multics_min}
\end{equation}
where the sparse signal $\bf S$ is a $Q\times L$ matrices and the received data $\bf Y$ has dimensions $M \times L$.} The equation is transformed into the equation (\ref{eq:multics_w}),(\ref{eq:multics_ws}).
\begin{equation}\label{eq:multics_w}
w_q^{k+1} = \frac{1}{\left|(\hat{s}_q^k)_{l_1}\right|+\xi}
\end{equation}
\begin{equation}
\left|w_q^{k+1}(s_q^{k+1})_{l_1}\right|=\left\{
\begin{array}{rcl}
\frac{(s_q^k)_{l_1}}{(s_q^k)_{l_1}+\xi}\approx 1, &    &\left|(s_q^k)_{l_1}\right|>0\\
0,&      &\left|(s_q^k)_{l_1}\right|=0
\end{array}
\right.
\label{eq:multics_ws}
\end{equation}

where $(s_q^k)_{l_1}$ is the $q^{th}$ element of the vectors ${\bf S}_{l_1}$ at the $k^{th}$ iteration. 
}

\section{Subspace-based compressive sensing algorithm}

 {In this paper, the broadband acoustic pulse emission combined to a wavelength that is small compared to water depth allows us to describe the waveguide propagation with the ray model. In the case of tank data, the propagation is dominated by reflections on the top and bottom waveguide boundaries while both refraction and reflections in shallow waters are observed with the ocean data used in the present study (Fig. 3c). This means that the $P$ largest eigenvalues of the spectral covariance matrix can be associated to the $P$ main eigenrays, which may not be the case in oceanic waveguides where modal dispersion is dominating.} We first compute the spectral matrix of the received signals as follows.
{\begin{equation}
	\hat{\mathbf{R}}=E \{ \mathbf{Y} \mathbf{Y^H}\}
%	=\begin{bmatrix}
%	E\{ Y_1Y^H_1\} & E\{ Y_1Y^H_2\} & \cdots & E\{ Y_1Y^H_M\}\\ 
%	E\{ Y_2Y^H_1\} & E\{ Y_2Y^H_2\} & \cdots & E\{ Y_2Y^H_M\}\\ 
% 	\vdots & \vdots &  & \vdots \\ 
%	E\{ Y_MY^H_1\} & E\{ Y_MY^H_2\} & \cdots & E\{ Y_MY^H_M\}
%	\end{bmatrix}
\end{equation}}
where $\hat{\mathbf{R}} \in \mathbb{R}^{M\times M}$. $E \{ \cdot \}$ denotes the expectation. The expectation is computed by a frequency-smoothing operation \cite{jiang2017raypath}. $\cdot ^\textbf{H}$ denotes the conjugate transpose. {In addition, the multiple raypaths are produced by the same source, and there exist correlations between the raypaths, which inevitably lead to rank deficiency, so we use frequency smoothing for preprocessing. For the wideband signal, two-sided correlation (TSC) matrices are designed at each frequency. Each transforms the spectral matrix at a certain frequency into the focusing spectral matrix at central frequency \cite{Valaee1995Wideband}.} 
Then, we apply an eigenvalue decomposition to the spectral matrix $\hat{\mathbf{R}}$. That is,
%{
%\begin{equation}
%\begin{split}
%	\hat{\mathbf{R}}=&\hat{\mathbf{R}}_s+\hat{\mathbf{R}}_n \\
%	                        %=&\mathbf{U} \mathbf{\Lambda} \mathbf{U}^H \\
%	                       % =& \sum_{k=1}^{M} \lambda_k \mathbf \mu_k  \mathbf \mu_k^H \\
%	 =& \sum_{k=1}^{P} \lambda_k \mathbf \mu_k \mathbf \mu_k^H  + \sum_{k=P+1}^{M} \lambda_k \mathbf \mu_k  \mathbf \mu_k^H 
%\end{split}
%\end{equation}}
{\begin{equation}
	\hat{\mathbf{R}}=\hat{\mathbf{R}}_s+\hat{\mathbf{R}}_n  = \sum_{k=1}^{P} \lambda_k \mathbf \mu_k \mathbf \mu_k^H  + \sum_{k=P+1}^{M} \lambda_k \mathbf \mu_k  \mathbf \mu_k^H 
\end{equation}}
%Eigenvalue decomposition can be applied to $\hat{\mathbf{R}}$ as
%\begin{equation}
%\begin{split}
%	\hat{\mathbf{R}}=&\mathbf{U} \mathbf{\Lambda} \mathbf{U}^H \\
%	 =& \sum_{k=1}^{M} \lambda_k \mu_k \mu_k^H \\
%	 =& \sum_{k=1}^{P} \lambda_k \mu_k \mu_k^H  + \sum_{k=P+1}^{M} \lambda_k \mu_k \mu_k^H 
%\end{split}
%\end{equation}
where $\lambda_1, \lambda_2, \cdots, \lambda_M$ are the eigenvalues of $\hat{\mathbf{R}}$. The eigenvalues are sequentially arranged as $\lambda_1 \geq \lambda_2 \geq \cdots \geq \lambda_M$, and 
%$\mathbf{\Lambda}=diag [\lambda_1, \lambda_2, \cdots,\lambda_M]$ denotes the diagonal matrix formed by the eigenvalues.
%Where $\mathbf{\Lambda}=\begin{bmatrix}
%\lambda_1 &0  &\cdots  &0 \\ 
%0 &\lambda_2  &  & \vdots \\ 
%\vdots &  &  \ddots & 0\\ 
%0 &  \cdots &  0 & \lambda_M
%\end{bmatrix}$ 
$\mathbf{\mu}_1, \cdots , \mathbf{\mu}_M$ denote the corresponding eigenvectors; 
%$\mathbf{U}$ is the unitary matrix with a size of {$M \times M$}. 
Therefore, the signal subspace is projected by the largest $P$ eigenvalues (We assume that the number of raypaths $P$ is known in this paper{\cite{Jiang2014Automatic}}.) and the corresponding eigenvectors.
\begin{equation}
	\hat{\mathbf{R}}_s=\sum_{k=1}^{P}\lambda_k \mathbf \mu_k \mathbf \mu_k^H
\end{equation}
 $\hat{\mathbf{R}_s}$ is rearranged into a long vector $\hat{\mathbf{r}_V} \in \mathbb{R}^{M^2 \times 1}$ by orderly stacking each row of $\hat{\mathbf{R}_s}$ behind the previous ones. 
%\begin{equation}
%	\hat{\mathbf{R}}_V=\begin{bmatrix}
%	R_{s,1,1} \\ 
%	\vdots \\ 
%	R_{s,1,M} \\ 
%	\vdots \\ 
%	R_{s,M,1}\\ 
%	\vdots \\ 
%    R_{s,M,M}
%\end{bmatrix}
%\end{equation}
That is, $\hat{\mathbf{r}}_V=[R_{s,1,1},R_{s,1,M}, R_{s,M,1}\cdots, R_{s,M,M}]^T$,
where $R_{s,i,j}$ denotes the $i^{th}$ row $j^{th}$ column element of $\hat{\mathbf{R}_s}$.

According to Eq. 3, we have
\begin{equation}
	E \{ \mathbf{YY}^H \} = E \{ \mathbf{GSS}^H \mathbf{G}^H \} + E \{ \mathbf{GSN}^H \} + E \{ \mathbf{NS}^H \mathbf{G}^H\} + E \{ \mathbf{NN}^H \}
\end{equation}
Assuming that the signals and the noise are uncorrelated, we obtain
\begin{equation}
	E \{ \mathbf{GSN}^H \} = E \{ \mathbf{NS}^H \mathbf{G}^H\} = \mathbf{0}
\end{equation}
and the signal subspace is denoted as follows:
\begin{equation}
	% \hat{\mathbf{R}}_s = E \{ \mathbf{GSS}^H \mathbf{G}^H \} + E \{ \mathbf{GSN}^H \} + E \{ \mathbf{NS}^H \mathbf{G}^H\}
	\hat{\mathbf{R}}_s = E \{ \mathbf{GSS}^H \mathbf{G}^H \}
\end{equation}

% Assuming the raypaths are incoherent (that is $E\{ S_i S_j^H \}=0, \forall i\neq j$), we have
In the context of OAT, {the raypaths are produced by the same source. {They may be de-correlated when they scatter at different points on the sea surface or cross different areas of the water column; however, there is still the possibility that the case of correlated or coherent raypaths exists.} To further improve separation resolution, it is necessary to consider and suppress the interferences produced by the correlations between the raypaths. Thus, the above equation can be rewritten as follows:
{\begin{equation}
\begin{split}
	\mathbf{GSS}^H \mathbf{G}^H  = 
	%& 
%	\begin{bmatrix}
%	 G_{11} & \cdots & G_{1N} \\ 
%	 \vdots & \ddots & \vdots \\ 
% 	 G_{M1} & \cdots & G_{MN} 
%	\end{bmatrix}
%	\begin{bmatrix}
% 	S_1S_1^H& 0 & \cdots & 0 \\ 
% 	0 & S_2S_2^H & \cdots & 0 \\ 
% 	\vdots & \vdots & \ddots & \vdots \\ 
% 	0 & 0 & \cdots & S_NS_N^H
%	\end{bmatrix}
%	\begin{bmatrix}
% 	G_{11}^H & \cdots & G_{M1}^H \\ 
% 	\vdots & \ddots & \vdots \\ 
% 	G_{1N}^H & \cdots & G_{MN}^H
%	\end{bmatrix} \\
%	+ & \begin{bmatrix}
%	 G_{11} & \cdots & G_{1N} \\ 
%	 \vdots & \ddots & \vdots \\ 
% 	 G_{M1} & \cdots & G_{MN} 
%	\end{bmatrix}
%	\begin{bmatrix}
% 	0 & S_1S_2^H & \cdots & S_1S_N^H \\ 
% 	S_2S_1^H & 0 & \cdots & S_2S_N^H \\ 
% 	\vdots & \vdots & \ddots & \vdots \\ 
% 	S_NS_1^H & S_NS_2^H & \cdots & 0
%	\end{bmatrix}
%	\begin{bmatrix}
% 	G_{11}^H & \cdots & G_{M1}^H \\ 
% 	\vdots & \ddots & \vdots \\ 
% 	G_{1N}^H & \cdots & G_{MN}^H
%	\end{bmatrix} \\
	=& \mathbf{D}_1 + \mathbf{D}_2
\end{split}
\end{equation}}
where \\
\begin{equation}
\begin{split}
	\mathbf{D}_1=\begin{bmatrix}
	 G_{11} & \cdots & G_{1N} \\ 
	 \vdots & \ddots & \vdots \\ 
 	 G_{M1} & \cdots & G_{MN} 
	\end{bmatrix}
	\begin{bmatrix}
 	S_1S_1^H& 0 & \cdots & 0 \\ 
 	0 & S_2S_2^H & \cdots & 0 \\ 
 	\vdots & \vdots & \ddots & \vdots \\ 
 	0 & 0 & \cdots & S_NS_N^H
	\end{bmatrix}
	\begin{bmatrix}
 	G_{11}^H & \cdots & G_{M1}^H \\ 
 	\vdots & \ddots & \vdots \\ 
 	G_{1N}^H & \cdots & G_{MN}^H
	\end{bmatrix} \\
\end{split}
\end{equation}

and \\
\begin{equation}
\begin{split}
	\mathbf{D}_2=\begin{bmatrix}
	 G_{11} & \cdots & G_{1N} \\ 
	 \vdots & \ddots & \vdots \\ 
 	 G_{M1} & \cdots & G_{MN} 
	\end{bmatrix}
	\begin{bmatrix}
 	0 & S_1S_2^H & \cdots & S_1S_N^H \\ 
 	S_2S_1^H & 0 & \cdots & S_2S_N^H \\ 
 	\vdots & \vdots & \ddots & \vdots \\ 
 	S_NS_1^H & S_NS_2^H & \cdots & 0
	\end{bmatrix}
	\begin{bmatrix}
 	G_{11}^H & \cdots & G_{M1}^H \\ 
 	\vdots & \ddots & \vdots \\ 
 	G_{1N}^H & \cdots & G_{MN}^H
	\end{bmatrix} \\
\end{split}
\end{equation}

Stacking each row of $\mathbf{D}_1$ below the previous ones, we rearrange $\mathbf{D}_1$ into a long vector $\mathbf{d_{1V}}$ and further reshape it as follows:
{\begin{equation}
\begin{split}
	\mathbf{d_{1V}}= 
	%&
%	            \begin{bmatrix}
%				S_1 S_1^H G_{11} G_{11}^H + \cdots S_N S_N^H G_{1N} G_{1P}^H \\ 
%				\vdots \\ 
%				S_1 S_1^H G_{11} G_{M1}^H + \cdots S_N S_N^H G_{1N} G_{MP}^H \\ 
%				\vdots \\ 
%				S_1 S_1^H G_{M1} G_{11}^H + \cdots S_N S_N^H G_{MN} G_{1P}^H \\ 
%				\vdots \\ 
%				S_1 S_1^H G_{M1} G_{M1}^H + \cdots S_N S_N^H G_{MN} G_{MP}^H
%				\end{bmatrix} \\
%			  = & 
%			    \begin{bmatrix}
% 				G_{11} G_{11}^H & G_{12} G_{12}^H & \cdots & G_{1N} G_{1N}^H \\ 
% 				G_{11} G_{21}^H & G_{12} G_{22}^H & \cdots & G_{1N} G_{2N}^H \\ 
% 				&  & \vdots & \\ 
% 				G_{M1} G_{M1}^H & G_{M2} G_{M2}^H & \cdots & G_{MN} G_{MN}^H
%				\end{bmatrix}
%				\begin{bmatrix}
%				S_1 S_1^H \\ 
%				S_2 S_2^H \\ 
%				\vdots \\ 
%				S_N S_N^H
%				\end{bmatrix}
			  = {\mathbf{G}}' \mathbf{S^{'}}
\end{split}
\end{equation}}
where 
$
	{\mathbf{G}}' = \begin{bmatrix}
 					G_{11} G_{11}^H & G_{12} G_{12}^H & \cdots & G_{1N} G_{1N}^H \\ 
 					G_{11} G_{21}^H & G_{12} G_{22}^H & \cdots & G_{1N} G_{2N}^H \\ 
 					&  & \vdots & \\ 
 					G_{M1} G_{M1}^H & G_{M2} G_{M2}^H & \cdots & G_{MN} G_{MN}^H
					\end{bmatrix}
					\in \mathbb{C}^{M^2 \times N}
$ and \\
$
	\mathbf{S^{'}}= \begin{bmatrix}
 				S_1 S_1^H & S_2 S_2^H & \cdots & S_N S_N^{H} 
				\end{bmatrix}^T \in \mathbb{R}^{N \times 1}
$.  $\cdot ^T$ denotes the matrix transpose.

{Similarly, we rearrange $\mathbf{D}_{2}$ into a long vector $\mathbf{{d}_{2V}}$ by stacking each row of $\mathbf{D}_2$ below the previous ones}, and $\mathbf{{d}_{2V}}$ will be set as a variable in the calculation with the convex optimization tools. 

Finally, we obtain
\begin{equation}
	\mathbf{r}_V={\mathbf{G}}' \mathbf{S^{'}} + \mathbf{{d}_{2V}}
\end{equation}
Based on CS theory, an accurate reconstruction is assured under two conditions: sparsity of the underlying signal and sufficient incoherence of the process that maps the underlying signals to the observations. In our case, $Q>>P$ assures sparsity, as there are only $P$ nonzero values. The incoherence of the process is assured, since the first column to the $M^{th}$ column of the sensing matrix ${\mathbf{G}}'$ constitute a Vandermonde matrix ({The element $a_{ij}$ at $i^{th}$ row and the  $j^{th}$ column of a Vandermonde matrix can be expressed as $a_{ij}=(a_i)^{j-1}$}.).  {If any two $P$ columns extracted from the sensing matrix are orthogonal, then the Restricted Isometry Property (RIP) condition is satisfied. At this time, the rank of the sub-matrix formed by the extracted arbitrary  two $P$ columns is $2P$. As a Vandermonde matrix satisfies the orthogonality of any  two $P$ columns mentioned above under certain $P${\cite{Jafarpour2011Deterministic,Cohen2009Compressed}}, Vandermonde matrixusually satisfies the RIP condition of compressed sensing.}

%Like the basic principle of CS-based algorithm, for sufficiently sparse signals and sensing matrices with sufficiently incoherent columns, 
%the proposed algorithm also exploits the spatial sparsity property based on the fact: when the raypaths of interest are spatially sparse, we only obtain a small number of direction of arrivals (DOAs) within the entire angle domain. The estimation of DOAs can be taken as a sparse-signal reconstruction problem. 
Thus, the sparse signal can then be reconstructed by solving the following {$L_1$} regularization optimization.
\begin{equation}
	\left\{\begin{matrix}
			\hat{\mathbf{S^{'}}} = \arg \min \left \| \mathbf{S^{'}} \right \|_1 , \text{ subject to } \mathbf{S^{'}} > 0 \\ 
			\left \| \mathbf{r}_V - {\mathbf{G}}' \mathbf{S^{'}} \right \|_2 \leq \delta, \delta > 0 
		   \end{matrix}\right.
\end{equation}
{We solved the above programming using available convex optimization tools, such as CVX \cite{Chen2001Atomic}.} The peak positions of $\hat{\mathbf{S^{'}}}$ represent the DOA estimations of the raypaths.
% Quality control editor: Journals often require both the manufacturer's name and location for specialized equipment and software. Please consider adding this information based on the journal’s guidelines.

{\section{Performance test}

%\subsection{Simulation}

In this section, we illustrate the performance of the subspace-based CS algorithm with simulation data in a multipath environment, a set of real data obtained in an ultrasonic waveguide{\cite{Roux2011Travel}} and, finally, a set of ocean shallow water data\cite{Roux2008The}. These tests are all performed in the same configuration, which is composed of a point source and a vertical array {of receivers}.  We used $150$, $300$ and $60$ snapshots in the simulations, tank experiments and at-sea experiments, respectively. The sound speed is uniform in the experiments, $1500$ m/s for the simulation, $1473$ m/s for tank data, $1509$ m/s for the real data. {We denote the wavelength as $\lambda$, the height of the water column as $H$, the interval between two adjacent sensors as $d$ and the distance between the point source and the reference sensor as $D$. Their specific values corresponding to each test are shown in Table 1. {In order to test the robustness of the subspace-based CS algorithm to the variation of sensor number and the interval between the two adjacent sensors, we chose vastly different values of $M$, and $d/\lambda$ for the three cases.}
}

%The point source is fixed at 50m under the water. 11 sensors are equidistantly fixed from 37.5 m to 62.5m under the water. The distance between the point source and the reference sensor of the vertical array is 2000 m. The depth of water is 100m. Five far-field signals are received by the array. Their theoretical DOAs are $-7.6^{\circ}$, $-3.8^{\circ}$, $0^{\circ}$, $3.8^{\circ}$ and $7.6^{\circ}$. The number of snapshots used in the simulations is {\color{red}200}. 

 MUSIC and reweighed CS beamforming \cite{Xenaki2014Compressive}
% $ L_1$-SVD \cite{malioutov2005sparse} and covariance-based compressive beamforming \cite{zhong2013compressive}, 
 are provided as comparative algorithms. Fig. 1 shows the received signals in simulations, tank experiments and at-sea experiments.  Figs. 2, 3 and 4 illustrate the {separation contrast results with simulation data} in a multipath environment, a set of real data obtained in an ultrasonic waveguide and, finally, a set of ocean shallow water data, respectively. {The SNR is defined as the ratio of signal power to the noise power in the frequency band of the signal for the array.  It is denoted by the equation  $ SNR = 10*lg (P_s / P_n)$, where $P_s$ and $P_n$ represent the power of signal and noise in the frequency band of the signal, respectively.} 
%  {\color{blue}The definition of the SNR in the paper is shown as follows. 
%\begin{equation}
%\centering
%\text{SNR}=10 \text{log} [ \frac{\sum_{p=1}^{P}|S_p(\nu)|^2}{\sum_{p=1}^{P}|N_p(\nu)|^2} ]
%\end{equation}}

Using different line types, Fig. 2(a) shows the separation results with simulation data when the SNR is equal to -5 dB, and the black crosses denote the theoretical positions. {The theoretical positions indicate the 
%Editor: Please ensure that the intended meaning has been maintained in the follow edit.
{correct arrival angles (or arrival times)}, 
which are computed using ray theory  \cite{Cornuelle1987Compressive,Skarsoulis2004Travel}.}
% Quality control editor: Abbreviations and acronyms (e.g., SNR) are typically defined the first time the term is used within the main text and then used throughout the remainder of the manuscript. Please consider adhering to this convention. The target journal may have a list of abbreviations that are considered common enough that they do not need to be defined. 
  It is clear that the subspace-based CS algorithm can successfully separate all the five {raypaths}, while the MUSIC algorithm totally fails in separating them. In addition, the reweighed CS algorithm detects four peaks, but it is difficult to identify them by their expected positions. When the SNR is equal to 0 dB, five raypaths are successfully separated with the subspace-based CS algorithm, and the corresponding results are shown in Fig. 2(b). In addition, the MUSIC algorithm does not resolve any raypath. The reweighed CS algorithm provides four peaks, and each of them is located between two theoretical values. Thus, it is also difficult to know the corresponding theoretical values of these peaks and whether these peaks are pseudo-peaks. 
% Quality control editor: Please ensure that the intended meaning has been maintained in the edits of the previous sentence. 
 Moreover, we test the performance of the subspace-based CS algorithm at different SNRs and compare its separation root-mean-square errors (RMSE) with those of other algorithms. {{We define the average RMSE for the direction of arrival ${\hat\theta_{p}}$ for the raypath $p$ by the quantity:}}

{{\begin{equation}
RMSE_{\theta_p}= \sqrt{\frac{1}{K_{ip}}\sum_{k=1}^{K_{ip}}|{\hat \theta_{p}}-\theta_{p}|^{2}},
\end{equation}}}

{Where $K_i$ is the number of the trials, which is equal to 20 for the simulation.  That is, the simulations are performed 20 times for different SNRs. Generally, the subspace-based CS algorithm is more robust to noise, as shown in Figs. 2(a) and (b). In Fig. 2(c), even for these raypaths, which can be separated by all three algorithms, the subspace-based CS algorithm achieves the minimum RMSEs.}
 
 Fig. 4 shows the separation contrast with the data obtained {in an ultrasonic waveguide immersed in a water tank}. {If the frequency of the signals is multiplied by a factor and the spatial distances, including both the one between the source and the receivers and the one between the adjacent receivers, are divided by the same factor, then the physical phenomena occurring in the environment remain the same. Thus, the small-scale experiment can reproduce the actual physical phenomena occurring in nature at a smaller scale inside the laboratory, which achieves a reduced cost and a totally controlled experiment. {In the present work,} a steel bar acts as the bottom, for which the {boundary conditions} are nearly perfect at the water-bottom interface, and a ~1.10 m long, 5.4 cm deep acoustic waveguide is constructed {\cite{Roux2014Inverting}}. Two coplanar 64-element vertical line arrays (VLAs) are placed, and {A broadband 1-{$\mu$s}-long ultrasonic pulse is  emitted by each source successively at the 3 MHz central frequency of the transducers \cite{Roux2011Travel}.} The transducer {dimensions} $0.75~mm \times12~mm$ {are} used to make the linear arrays omnidirectional in the plane defined by the source-receiver arrays, {and the beams are collimated in the plane perpendicular to the waveguide axis. {We chose five point-to-array configurations (one emitted source and 31 receivers ) and we described their specific positions for the benchmark configuration in Table I.}} {{Five eigenrays are expected {between the source-receiver arrays} with this experimental configuration.} To further test the performance of the subspace-based algorithm, we study statistically the outputs of the beamforming algorithms using a set of sources around the actual source [the two above and the two below in the source array (SA)] and sets of subarrays among the vertical received array that are close to the center of the receiver array (RA). The parameters of the benchmark configuration are listed in the second line of Table I. In addition,  the noise (SNR=-3 dB) has been added to the received signal for the ultrasound scenario. 
 %We finally choose nine sets of data in total to test the performance in this paper.
% The parameters are listed in Table 2.
 The same processing using the subspace-based CS algorithm, the CS beamforming algorithm and the MUSIC algorithm was performed {to produce the beamformer outputs.}
 In Fig. 3(a), the average {beamformer outputs are shown. The purple line corresponds to the average output of the
 CS beamforming algorithm; the blue line with circles corresponds to that of the subspace-based CS algorithm; and the yellow line with very short vertical lines indicates that of the MUSIC algorithm.}
  Clearly, the subspace-based CS algorithm correctly separates all the five raypaths with minimal bias {compared to} the theoretical position. At the same time, the MUSIC algorithm fails in separating three of the five raypaths, and the CS beamforming algorithm gives additional pseudo-peaks. In Fig. 3(b), only the peaks of these spectra are retained.  {The purple triangles denote the peaks of the CS beamforming algorithm; the blue circles represent those of the subspace-based CS algorithm; and the yellow squares correspond to those of the MUSIC algorithm}. It can be seen that the subspace-based algorithm shows the smallest dispersion to the theoretical values of the DOAs, which means that it is more {robust to slight changes in the experimental configuration} than the other two algorithms. 
% The result of $L_1$-SVD gives a pseudo peak between $-2^{\circ}$ and $0^{\circ}$. Similarly, a pseudo peak between $-4^{\circ}$ and $-2^{\circ}$ is detected by the covariance-based CS algorithm.

{Fig. 3 shows the contrasts of separation results with five data {sets} obtained in an at-sea experiment \cite{Roux2013Analyzing}. {The FAF05 (Focused Acoustic Field 2005) experiment was conducted in July 2005, to the north of Elba Island, Italy with repetitive data collection over more than eight consecutive hours  between two source-receiver vertical arrays that were separated by a distance 4.071 km in a 123 m deep waveguide.} The {FAF05} at-sea experimental setup is similar to that of the small-scale ultrasonic experiment{, although at a much larger scale}. The configuration is composed of two equally spaced VLAs. The source array (SA) has 29 transducers spanning 78 m in 120 m water, and the receiver array (RA) has 32 hydrophones covering 62 m. 
% The distance between the two VLAs is 4.071 km. 
 The central frequency of the transducers is 3.2 kHz, with 1 kHz bandwidth. {The source signals were 200-ms linear frequency modulated chirps that were compressed after reception to their pulse equivalent by cross-correlation. This provided 40 dB of signal-to-noise ratio for reception with power-limited transmission.}
% {\color{blue}The 200-ms linear frequency modulated chirps were used as the source signals. Each transducer transmitted the source signal sequentially and the time interval between transmissions is 250 ms which is greater than the channel dispersion time. On Julian Day 197, the waveguide transfer function is completely acquired in 7 s. The acquisition is repeated every 20 s. The total received signal for 8.5 h is recorded to monitor fluctuations of the oceanic waveguide.\cite{Roux2013Analyzing}} 
%{ {\color{blue}Not clear : just say you use five different subarrays that will provide different experimental configurations with arrivals angles that are close to each other... }
{Similar to the case of the tests with tank data, we use five different subarrays that will provide different experimental configurations with arrival angles that are close to each other. Fig. 4(c) shows the five eigenrays (including one surface-reflected ray and four refracted rays) propagating between two elements of the SA and the RA\cite{Roux2013Analyzing}. The eigenrays are either refracted at thermoclines or reflected at the air-water interface. {All of} the eigenrays {that} interact with the bottom and the surface-reflected eigenray {have} lower amplitudes than the refracted eigenrays because of reflection loss.}
%we chose five sets of data through removing the source around the reference one (the two above and the two below in the SA) and combining them with sets of subarrays among the vertical received array that are close to the center of the receiver array (RA). The parameters of the benchmark configuration are listed in the third line of the Table 1.
%To test the performance of the double-4-smoothing MUSICAL algorithm, the contrast experiments were made using a set of ocean data obtained in an at-sea experiment performed in July 2005 north of Elba Island, Italy \cite{Roux2013Analyzing}. 
%The parameters used in the tests are shown in table 3. 
%The reference source and the reference receiver are 4.701 km apart and located respectively at  90m under the surface. {\color{blue}non informative. Instead, say one sentence on the different types of work that were performed / published with this data set. }
{The FAF05 data set has been used to develop a coordinated source-receiver array processing procedure and illustrate its effectiveness in an example of tracking raylike arrivals in a fluctuating ocean environment \cite{Roux2008The}.
In addition, using these data, Roux et el. \cite{Roux2013Analyzing} analyzed the {time-dependent} sound speed fluctuations in shallow water from group-velocity versus phase-velocity data representation.} 
%Containing one surface-reflected ray and four refracted rays, a total of five rays propagate between two spots located at 90m beneath the surface where both centers of SA and RA lie. All the five raypaths consisting in the data are sketched in Fig (b).
 {The subspace-based CS algorithm, the CS beamforming algorithm and the MUSIC algorithm are applied to the five sets of data. We calculate the expected DOAs based on the propagation of the raypaths. The average spectra for the five sets of data, shown in Fig. 4(a), are used to statistically study the performance of each algorithm. 
% Editor: The previous sentence was restructured to improve its clarity. Please ensure that the intended meaning has been maintained. 
 The subspace-based CS algorithm provides five peaks around the expected {theoretical DOAs}, while the CS beamforming algorithm gives multiple peaks with relatively large deviation. In addition, we only keep the peaks for each algorithm and plot them in Fig. 4(b). It can be seen that there is no dispersion in the results of the subspace-based CS algorithms and that there is much more dispersion with those of the CS beamforming algorithm. Additionally, although the results of the MUSIC algorithm gather together around the expected values, it fails in detecting some of the raypaths for {some subarray configurations}.}}
 {Based on these results, we can conclude that the subspace-based CS algorithm provides more reliable results with higher {angular} resolution.}}
 
%  \begin{table}
%\begin{center} \begin{tabular}{ c | c | c|c| c |c |c|c|c|} \hline & $M$ & $d (m)$ & $D (m)$ & $H$ (m)& $\lambda (m)$ &$d/\lambda$&$D/H$&D/$\lambda$\\ \hline \hline Simulation & 11 &   $2.5$& $2\times 10^{3}$ & 100 & 1&2.5&20& $2\times 10^{3}$ \\ \hline \hline
% Small-scale experiment  & 31& $1.5\times 10^{-3}$ &$1$  & $5.5\times 10^{-2}$ & $1.473\times 10^{-3}$& 1.02&18.18&678.89\\ \hline\hline Ocean data  & 5 &2 & $4.701\times 10^{3}$ &120 &  $0.43$&4.64&39.175 & $1.0914\times 10^{4}$ \\ 
%\hline  
% \end{tabular} \end{center}
%\caption{The configuration parameters of the tests}
%\end{table}
 
 \newcommand{\tabincell}[2]{\begin{tabular}{@{}#1@{}}#2\end{tabular}}

\begin{table}
\begin{center} 
\begin{tabular}{ c | c | c|c| c |c |c|c|c|} 
\hline & $M$ & $d~(m)$ & $D~(m)$ & $H$ (m)& $\lambda~(m)$ &$d/\lambda$&$D/H$&D/$\lambda$\\ \hline 
\hline Simulation & 11 &   $2.5$& $2\times 10^{3}$ & 100 & 1&2.5&20& $2\times 10^{3}$ \\ \hline \hline
 \tabincell{c}{Small-scale \\experiment}  & 31& \tabincell{c}{$1.5\times$\\ $ 10^{-3}$} &$1$  & \tabincell{c}{$5.5\times$ \\$10^{-2}$} & \tabincell{c}{$1.473$\\ $\times 10^{-3}$}& 1.02&18.18&678.89\\ \hline\hline Ocean data  & 5 &2 & $4.701\times 10^{3}$ &120 &  $0.43$&4.64&39.175 & \tabincell{c}{$1.0914$ \\ $\times 10^{4}$} \\
\hline
 \end{tabular} \end{center}
\caption{The configuration parameters of the tests, where $M$ is the number of sensors, $d$ is the interval between two adjacent sensors, $D$ is the distance between the point source and the reference sensors, $H$ is the height of the water column and $\lambda$ is the wavelength.}
\end{table}

\begin{figure}[tbhp!]
\begin{center}
 % \subfigure[]{\includegraphics[width=9.6cm,height=6.8cm]{simu_emited.jpeg}\label{FIGU_3a}}\hspace{+0.1cm}
  \subfigure[]{\includegraphics[width=9cm, height=6cm]{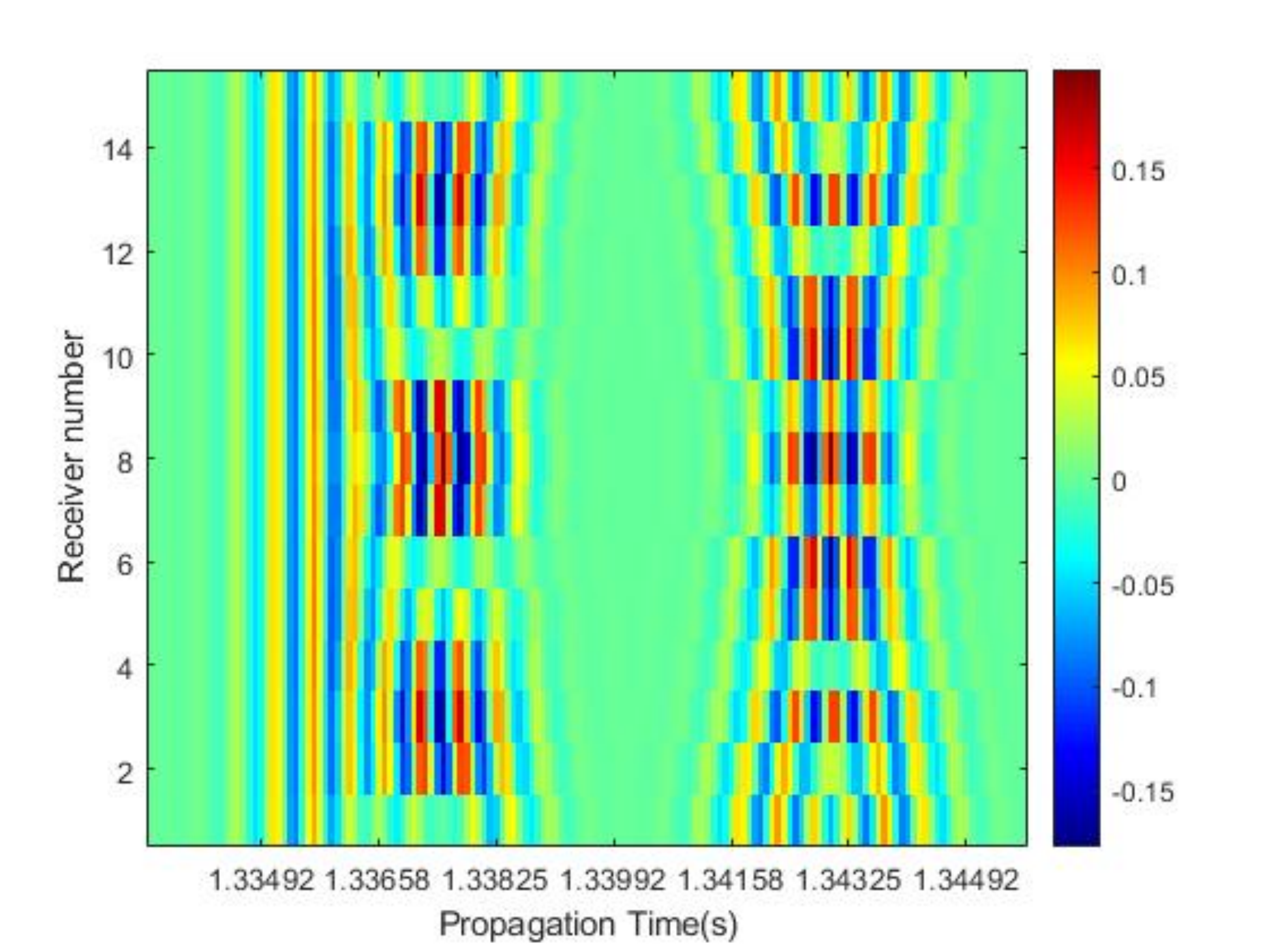}\label{FIGU_3b}}\hspace{+0.1cm}
    \subfigure[]{\includegraphics[width=9cm, height=6cm]{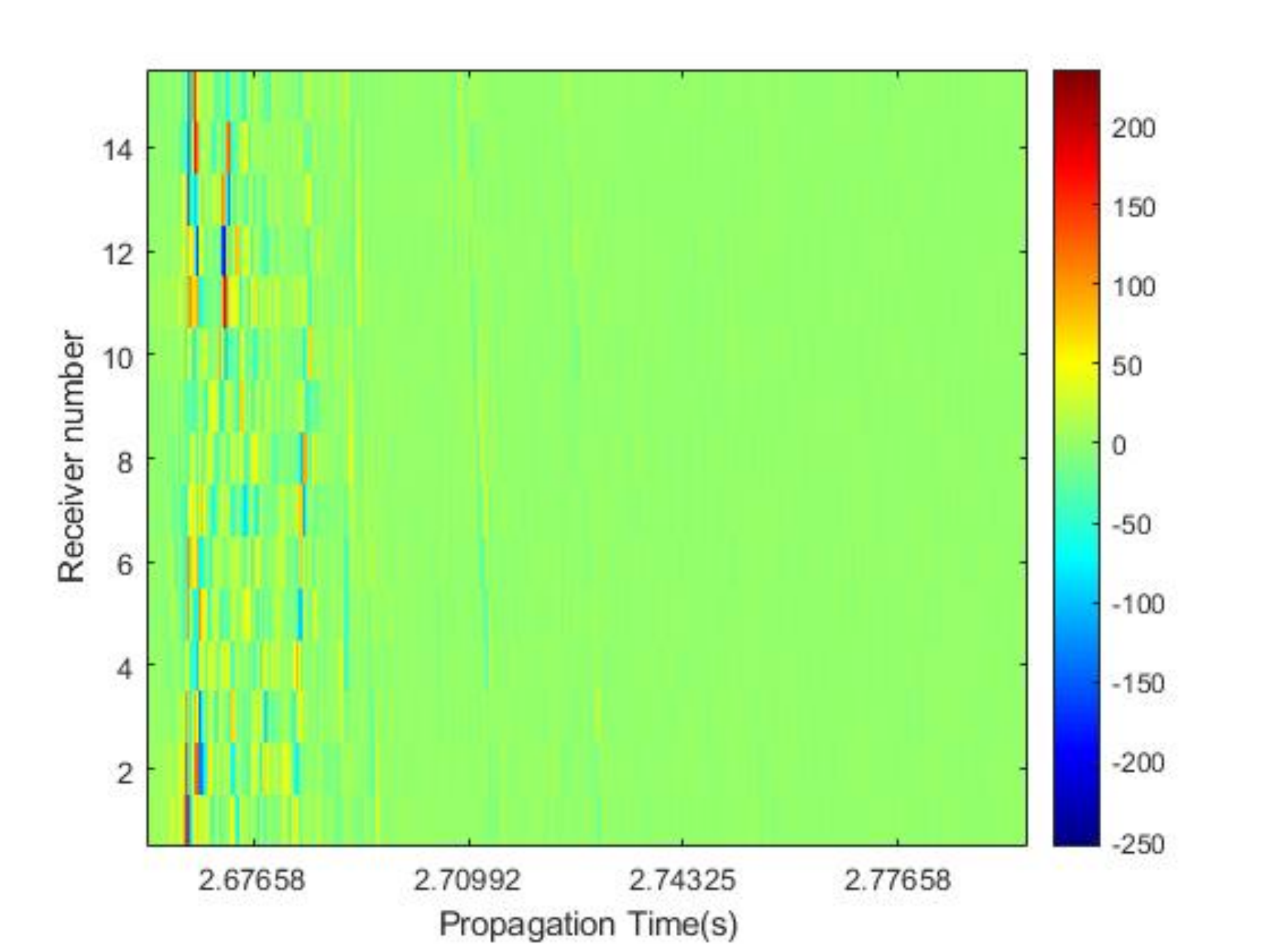}\label{FIGU_3b}}\hspace{+0.1cm}
 \subfigure[]{\includegraphics[width=9cm, height=6cm]{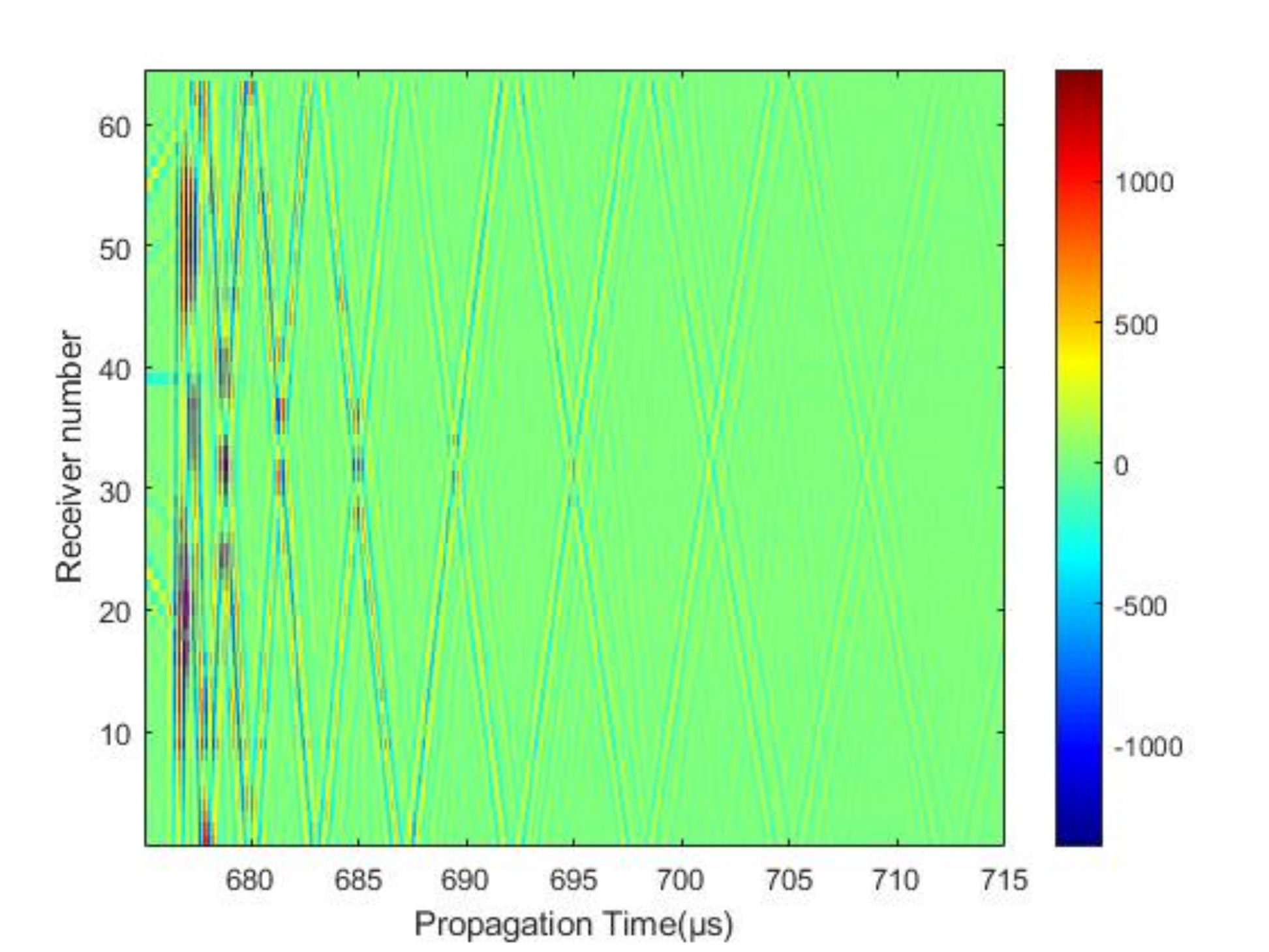}\label{FIGU_3b}}\hspace{+0.1cm}
 % \subfigure[ ]{\includegraphics[width=7.5cm,height=5.5cm]{eig_epn_4s1}\label{FIGU_c2}}\hspace{+0.1cm}
\end{center}
  \vspace{-0.4cm}
\caption {\label{FIGU1}The received signals: (a) simulations; (b) tank experiments; (c) at-sea experiments.}
\end{figure}

 \begin{figure}[h!]
\centering    
\includegraphics[width=188mm, height=158mm]{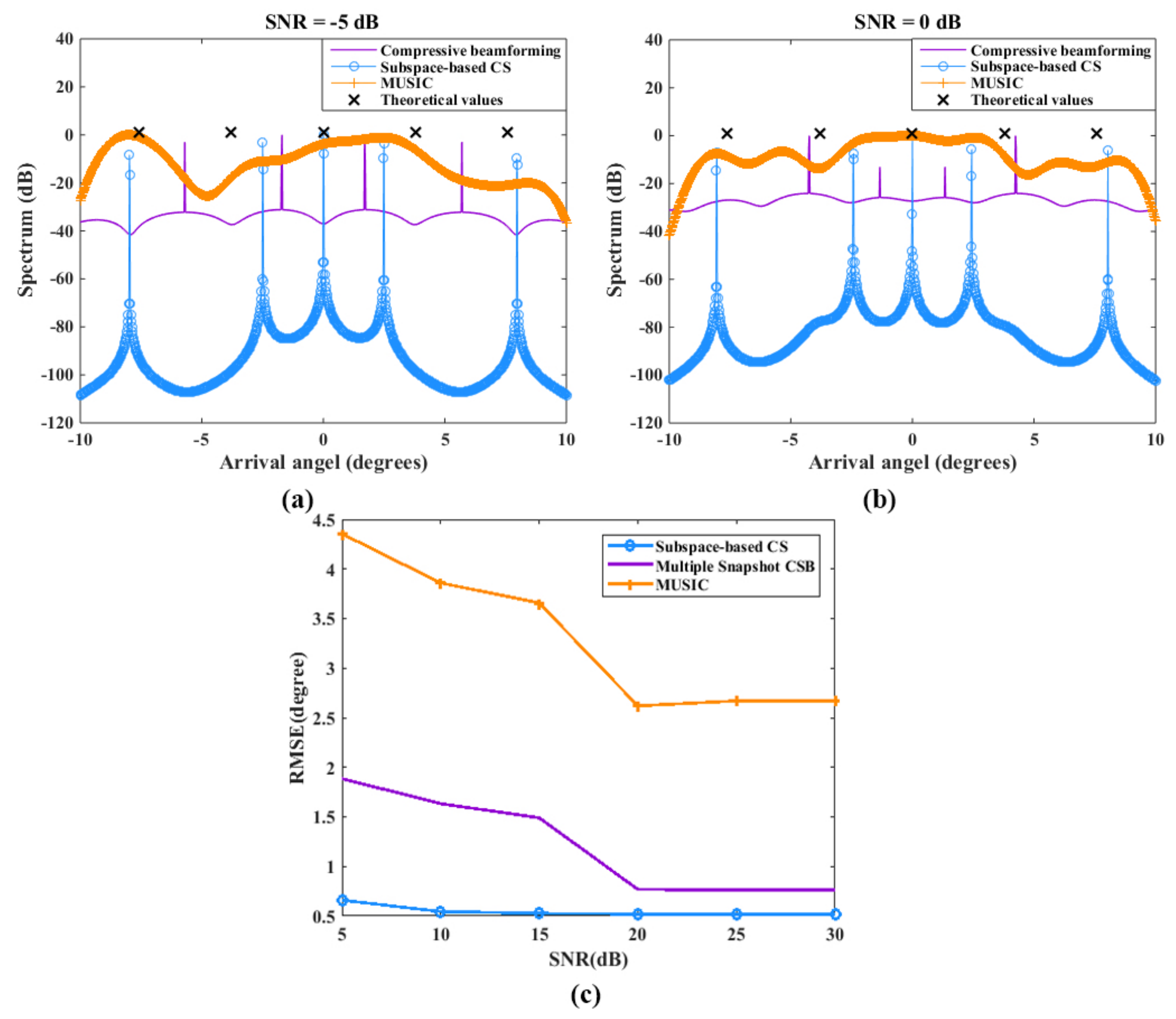}
\caption{\label{FIGU_1}Separation results comparison of different algorithms using simulation data. (a) SNR= -5 dB; (b) SNR= 0 dB; (c) RMSE comparison }
\end{figure}

%\begin{table}
%\begin{center} \begin{tabular}{ c | c | c|c| c |c } \hline & $M$ & $z_s (mm)$ & $z_{r_1} \sim z_{r_M} (mm)$ & $d$ (mm)& $D (mm)$\\ \hline \hline (a) &  \multirow{9}{*}{31}&  \multirow{3}{*} {26.375} & $3.125 \sim 48.125$ & \multirow{9}{*}{1.5} & \multirow{9}{*}{$1\times 10^{3}$}  \\ (b)  & &  &$4.625 \sim 49.625$  & & \\  (c) &  &  & 6.125 $\sim$ 51.125 & & \\  \cline{3-4} (d)  &  &  \multirow{3}{*} {27.125} & 3.125 $\sim$ 48.125 &  &  \\  (e) &   &  & 4.625 $\sim$ 49.625 &  & \\ (f)  &  &  & 6.125 $\sim$ 51.125 &  &  \\   \cline{3-4} (g)  &   &\multirow{3}{*} {27.875}  & 3.125 $\sim$ 48.125 &  &  \\  (h)  &  &  & 4.625$\sim$ 49.625 &  &  \\ (i)  &  &  & 6.125 $\sim$ 51.125 &  &  \\ \hline \hline \end{tabular} \end{center}
%\caption{The configuration parameters of  small-scale experiment (nine sets)}
%\end{table}

%\begin{table}
%\begin{center} 
%\begin{tabular}{ c | c | c } \hline & $\nu_c$ (Hz)  & $ N$  \\ 
%\hline \hline Simulation & $1.5 \times10^{3}$ & 150 \\ 
%\hline \hline Small-scale experiment  &1.2 $ \times10^{6}$  & 300 \\ 
%\hline\hline Ocean data  & $3.5 \times10^{3}$ & 537  \\ 
%\hline   
%\end{tabular} 
%\end{center}
%\caption{{The test parameters used in the simulations, the small-scale experiment and the at-sea experiment.}}
%\end{table}

\begin{figure}[h!]
\begin{center}
	%{\includegraphics[width=168mm,height=138mm]{fig2(1).png}\label{FIGU_1}}
%	\subfigure[]{\label{fig:a}\includegraphics[width=95mm]{tank-average-curve}}\vspace{-0.2cm}
%	\subfigure[]{\label{fig:b}\includegraphics[width=95mm]{peaks-tank}}\vspace{-0.2cm}
%\subfigure[]{\label{fig:a}\includegraphics[width=125mm]{fig3bf2}}
\includegraphics[width=138mm,height=188mm]{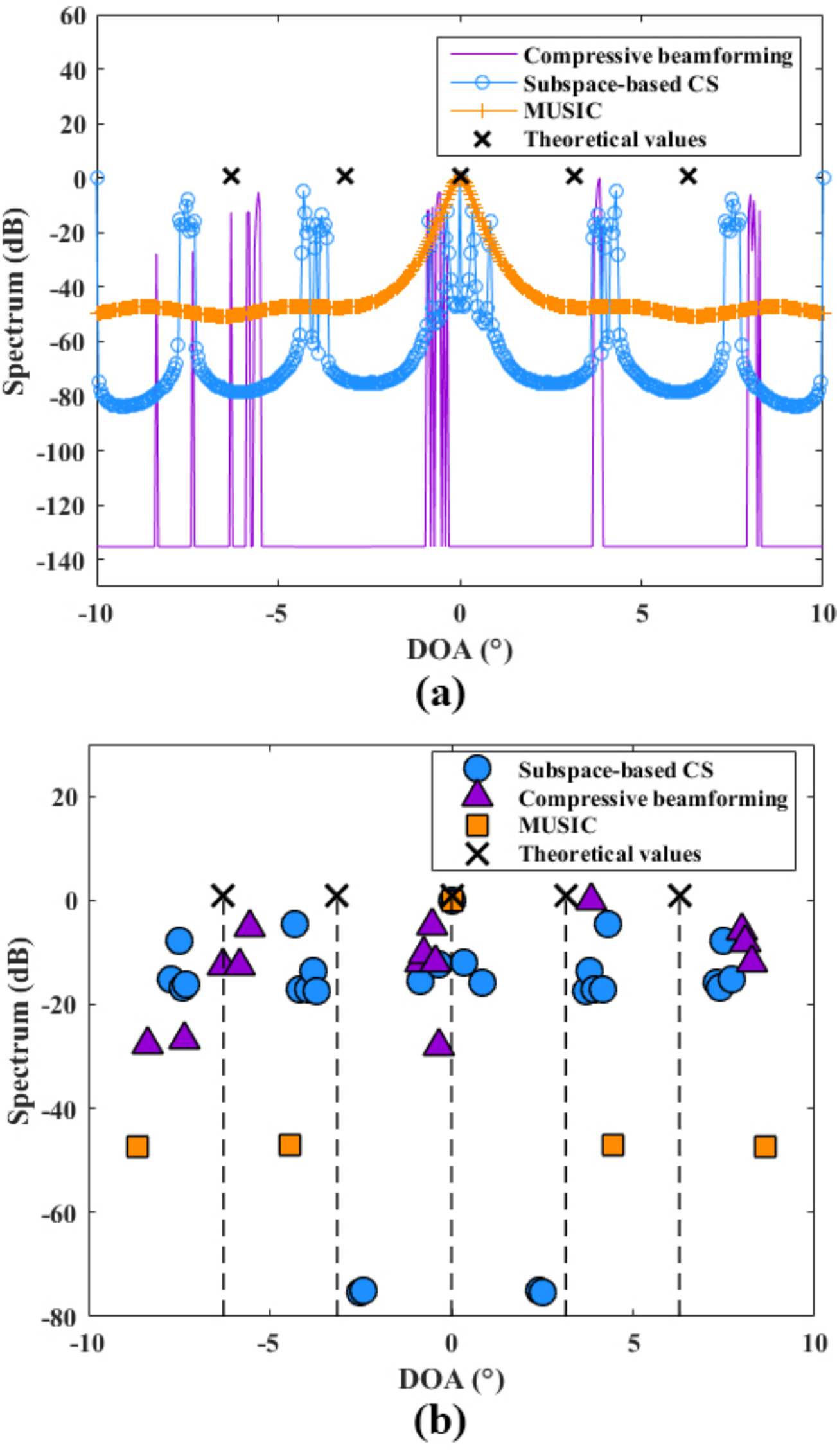}\label{FIGU_2}
\end{center}
\vspace{-1.5cm}
\caption {(color online) Separation results comparison of different algorithms in the case of data obtained in a tank (nine sets in total). A line type (or a mark) corresponds to a result of an algorithm. The black crosses denote the theoretical values, and the benchmarks are plotted with dashed lines. (a) Average; (b) Angular peaks.}
\label{FIGU1}
\end{figure}

%\begin{table}
%\begin{center} \begin{tabular}{ c | c | c|c| c |c } \hline & $M$ & $z_s (m)$ & $z_{r_1} \sim z_{r_M} (m)$ & $d$ (m)& $D (m)$\\ \hline \hline (a) &   5 &  93.0980& $92 \sim 100$ & 2& $4.701\times 10^{3}$ \\ \hline \hline (b)  &5 & 93.098 &$88 \sim 96$  & 2 & $4.701\times 10^{3}$ \\ \hline\hline (c) & 7 & 95.8804 & 84 $\sim$ 96 &2 &$4.701\times 10^{3}$  \\ \hline   \hline (d)  & 7 & 93.098 & 84 $\sim$ 96 & 2 & $4.701\times 10^{3}$  \\ \hline \hline (e) & 9  & 95.8840 & 82 $\sim$ 98 & 2 & $4.701\times 10^{3}$ \\ \hline \end{tabular} \end{center}
%\caption{The configuration parameters of the at-sea experiment (five sets)}
%\end{table}

\begin{figure}[h!]
\begin{center}
	%\centering 
\vspace{-0.2cm}
%\subfigure[]{\label{fig:a}\includegraphics[width=80mm]{oceandata-average-curve}}
%\subfigure[]{\label{fig:b}\includegraphics[width=80mm]{peaks_ocean_data.png}}\vspace{-0.2cm}
%\subfigure[]{\label{fig:a}\includegraphics[width=88mm]{fig3af3}}
\includegraphics[width=188mm,height=158mm]{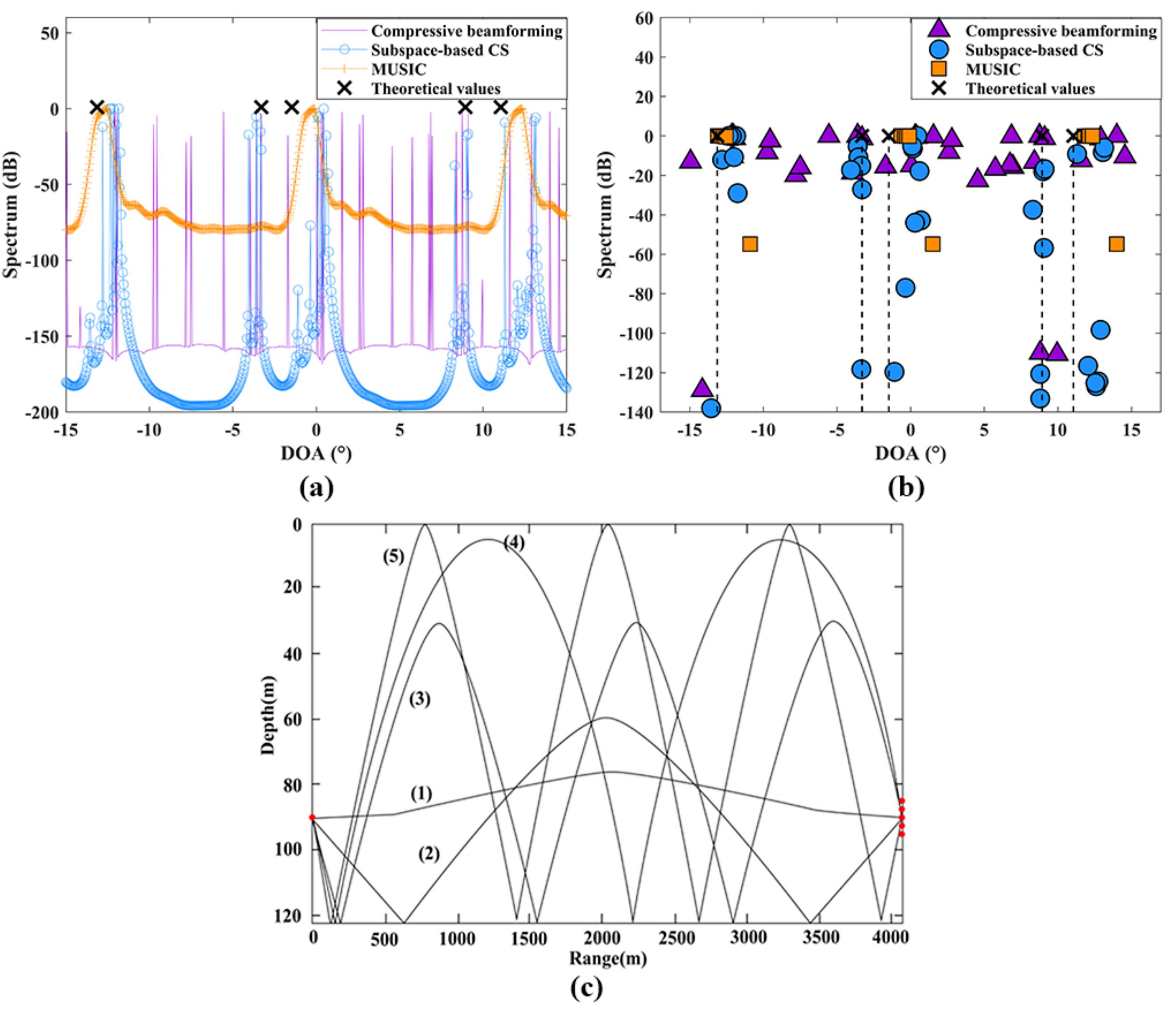}\label{FIGU_3}
%{\includegraphics[width=168mm,height=118mm]{fig3(1).png}\label{FIGU_3}}
\end{center}
\vspace{-1cm}
\caption{\label{FIGU1} Separation results comparison of the different algorithms in the case of ocean data obtained during the FAF05 at-sea experiment (five sets in total). A linetype (or a mark) corresponds to a result of an algorithm. The black crosses denote the theoretical values, and the benchmarks are plotted with dashed lines. (a) Average; (b) Spectrum peaks; (c) Set of eigenrays propagating between two elements of the SA and the RA (at a depth of 90 m). }

\end{figure}

\section{Conclusions}

In this paper, a subspace-based CS algorithm is proposed for separating raypaths, which introduces the subspace-based algorithm into a CS framework. {Compared to MUSIC and CS beamforming, the proposed algorithm achieved a better performance in terms of resolution and robustness to ocean fluctuations, especially when in a strongly noisy environment.}

%  End of title page for Preprint option --------------------------------- %

%% See preprint.tex/.pdf or reprint.tex/.pdf for many examples

%% before appendix (optional) and bibliography:
\begin{acknowledgments}
The Project Supported by the National Natural Science Foundation of China (No: 61871124 and 61876037), The national defense Pre-Research foundation of China, by the fund of Science and Technology on Sonar Laboratory (No: 6142109KF201806), by the Stable Supporting Fund of Acoustic Science and Technology Laboratory (No: JCKYS2019604SSJSSO12). The Focused Acoustic Forecasting experiment (FAF05) was performed in collaborative experiments with the NATO Underwater Research Centre (NURC), La Spezia, Italy, with Mark Stevenson as Chief Scientist. Scientists who contributed to these experiments include Tuncay Akal, W. A. Kuperman, W. H. Hodgkiss, H. C. Song, B. D. Cornuelle, Piero Boni, Piero Guerrini, other NURC staff, and the officers and crew of the RV Alliance.
\end{acknowledgments}

% -------------------------------------------------------------------------------------------------------------------
%   Appendix  (optional)

%\appendix
%\section{Appendix title}

%If only one appendix, please use
%\appendix*
%\section{Appendix title}

%=======================================================
%IMPORTANT

%Use \bibliography{<name of your .bib file>}+
%to make your bibliography with BibTeX. 

%Once you have used BibTeX you
%should open the resulting .bbl file and cut and paste the entire contents 
%into the end of your article.
 
%\bibliographystyle{ieeetr}
 % \bibliography{mybibtex}

%=======================================================

\end{document}